\documentclass[final,3p,times,mathpazo,twocolumn,fixfloat,sort&compress]{elsarticle}

\usepackage{graphicx}
\usepackage{subfig}

\usepackage{multirow}
\usepackage{amsmath}
\usepackage{comment}

\usepackage{tabularx,ragged2e}%

\newcolumntype{Y}{>{\centering\arraybackslash}X}

\usepackage{lineno}

\newcommand{\beq}{\begin{eqnarray}}
\newcommand{\eeq}{\end{eqnarray}}
\newcommand{\be}{\begin{eqnarray*}}
\newcommand{\ee}{\end{eqnarray*}}
\newcommand{\nn}{\nonumber}

\newcommand{\ep}{\varepsilon}

\newcommand{\vect}[1]{\overset{\to}{#1}}

\newcommand{\vectc}[1]{\vect{#1}\,^2}
\newcommand{\eqs}[1]{\begin{equation} \begin{split} #1\end{split} \end{equation} }

\newcommand{\ie}{{\it i.e.}}
\newcommand{\eg}{{\it e.g.}}

\newcommand{\sqrtsnn}{\sqrt{s_{NN}}}

\newcommand{\Lumi}{\mathcal{L}}

\newcommand{\lsim}{\mathrel{\vcenter
{\hbox{$<$}\nointerlineskip\hbox{$\sim$}}}}

\newcommand{\gsim}{\mathrel{\vcenter
{\hbox{$>$}\nointerlineskip\hbox{$\sim$}}}}

\newcommand{\ce}[1]{Eq.~\eqref{#1}}

\newcommand{\cf}[1]{{Fig.~\ref{#1}}}
\newcommand{\ct}[1]{{Tab.~\ref{#1}}}

\def\lsim{\raise0.3ex\hbox{$<$\kern-0.75em\raise-1.1ex\hbox{$\sim$}}}
\def\gsim{\raise0.3ex\hbox{$>$\kern-0.75em\raise-1.1ex\hbox{$\sim$}}}

\def\pp   {$pp$}
\def\pA   {$pA$}

\def\beq     {\begin{equation}}
\def\eeq     {\end{equation}}

\def\pp   {$pp$}
\def\pA   {$pA$}

\def\beq     {\begin{equation}}
\def\eeq     {\end{equation}}

\usepackage{ifpdf}
\ifpdf
\usepackage[pdftex]{hyperref}
\else
\usepackage[hypertex]{hyperref}
\fi

\hypersetup{
  pdftitle={},%
  pdfauthor={J.P. Lansberg,L. Massacrier, L. Szymanowski,J. Wagner},%
  pdfsubject={},%
  pdfkeywords={Quarkonium Production, Fixed-Target experiment, Large Hadron Collider, GPD, STSA},%
  pdfstartview={},%
  bookmarksopen=true, breaklinks=true, debug=true, %
  colorlinks=true, linkcolor=red, citecolor=blue, urlcolor=blue
}

\journal{Physics Letters B}

\begin{document}
\begin{frontmatter}

\title{Single-Transverse-Spin Asymmetries in Exclusive Photo-production of $J/\psi$ in Ultra-Peripheral Collisions in the Fixed-Target Mode at the LHC and in the Collider Mode at RHIC}

\author[a]{J.P.~Lansberg}
\author[a]{L.~Massacrier}
\author[b]{L.~Szymanowski}
\author[b]{J.~Wagner }
\address[a]{IPNO, CNRS-IN2P3, Univ. Paris-Sud, Universit\'e Paris-Saclay, 
91406 Orsay Cedex, France}
\address[b]{National Centre for Nuclear Research (NCBJ), Ho\.{z}a 69, 00-681, Warsaw, Poland}

\begin{abstract}
{\small We investigate the potentialities offered by the study of $J/\psi$ exclusive photo-production in ultra-peripheral collisions at a 
fixed-target experiment using the proton and lead LHC beams (generically denoted as AFTER@LHC) on hydrogen targets and at RHIC in the collider mode. We compare the expected counting rates in both set-ups. Studying Single-Transverse-Spin Asymmetries ($A_N$) in such a process provides a direct path to  the proton Generalised Parton Distribution (GPD) $E_g(x,\xi,t)$. We evaluate the expected precision on $A_N$ for realistic conditions with the LHCb detector in $p$H$^\uparrow$ and PbH$^\uparrow$ collisions. We also discuss prospects with polarised deuterium and
helium targets in the case of AFTER@LHC.}
\end{abstract}
\end{frontmatter}

\section{Introduction}
\label{intro}

The exclusive photo-/lepto-production of vector quarkonia, via $\gamma^{\star} p \to V p$, 
in the Bjorken limit is known to be a powerful
tool to probe the tri-dimensional gluon content of the proton. If, in addition,
it is studied on transversally polarised proton, it provides a direct access
to the orbital angular momentum carried by the gluons, $L_g$, which remains
unmeasured. 

The first attempt to perform this exclusive measurement was recently carried out by
the COMPASS collaboration~\cite{Matousek:2017yhq} in $J/\psi$ muo-production on a transversally 
polarised NH$_3$ target at $\sqrt{s}=17$ GeV in the limit where the $J/\psi$ takes the whole photon
momentum. Studies at higher energies will only be possible at a possible future EIC~\cite{Accardi:2012qut}. 

Beside lepton-induced reactions, the same sub process can also be accessed in proton-proton and
nucleus-proton by selecting collisions where a quasi-real photon is emitted by one proton 
or one nucleus. Such collisions are known as Ultra-Peripheral Collisions (UPC) and are routinely
studied in nucleus-nucleus collisions at RHIC~\cite{Adler:2002sc,Adams:2004rz,Afanasiev:2009hy} and the LHC~\cite{Abelev:2012ba,Abbas:2013oua,Adam:2015sia,Adam:2015gsa}. At the LHC, 
they are also studied in proton-nucleus collisions~\cite{TheALICE:2014dwa}. Along the same lines, 
exclusive proton-proton scatterings can occur via a photon emission from one proton~\cite{Aaij:2013jxj,Aaij:2014iea,Aaij:2015kea}. 

In~\cite{Lansberg:2015kha}, we have shown that UPCs can be studied in the fixed-target
mode at the LHC beams (such a mode will generically be referred to as AFTER@LHC~\cite{Hadjidakis:2018ifr,Brodsky:2012vg} in what follows) up to $\sqrt{s}= 40$ GeV (see also~\cite{Goncalves:2018htp,Massacrier:2017lib,Goncalves:2015hra}). In particular, pseudo-scalar
quarkonium or exclusive photo-production of a dilepton can be studied to measure the
quark GPDs. In this Letter, we demonstrate that  $\gamma^{\star} p^\uparrow \to J/\psi p$
can be accessed at AFTER@LHC via Single Transverse Spin Asymmetries (STSA or $A_N$) in UPCs 
in the same way that it can be accessed at RHIC
and can be used to put constrain on the GPDs $E_g(x,\xi,t)$.

The structure of this Letter is as follows. In section 2, 
we recall the main characteristics of the UPCs and the corresponding
photon fluxes in the fixed-target mode using the LHC beams. 
In section 3, we evaluate the expected cross sections for $J/\psi$  exclusive production
and the corresponding counting rates
both for AFTER@LHC and for RHIC based on {\sc Starlight}~\cite{Klein:2016yzr}. In section 4, we extend the 
discussion in terms of GPDs and show how STSAs allow one to access the GPD $E_g(x,\xi,t)$ and present the expected STSA magnitudes for AFTER@LHCb, namely with the LHCb detector is used.
Finally, we present our conclusions and  outlook for light nuclei.

\section{Ultra-peripheral collisions in the fixed-target mode at the LHC and in the collider mode at RHIC}

Charged hadrons moving at relativistic speed travel along electromagnetic fields 
which can be employed as quasi-real-photon beams. In the ultra-relativistic domain, 
the energy of these photons is such that they can trigger the production of hard dileptons, 
charmonia and even bottomonia, like at lepton-proton colliders.

The energy spectrum of these photons is usually computed in the Equivalent Photon Approximation (EPA) 
(see \eg~\cite{Budnev:1974de,Baltz:2007kq}). It depends on the boost between the charged hadron and the observer as well as on the impact
parameter $b$. In particular, the flux as function of the photon momentum $k$, of $b$ and $\gamma$ (that is the Lorentz factor of the hadron -- or nucleus -- in the frame where $k$ is measured)
reads~
\begin{equation}
\frac{dn}{dkd^2b} = \frac{Z^2\alpha_{\rm em} \omega(b,k)^2}{\pi^2 k b^2}
\left[K_1^2(\omega(b,k))+\frac{1}{\gamma^2}K_0^2(\omega(b,k)) 
\right],\label{eq:dndkd2b}
\end{equation}
where $\alpha_{\rm em}$ is the QED coupling, $Z$ is the emitter charge, $\omega(b,k)=kb /\gamma$ and
$K_{1,2}$ are modified Bessel functions of the second kind.  $b$ cannot be smaller than the hadron radius $R$, hence
the consideration of UPCs. If $b<R$, the probability for hadronic interactions may be higher than 
the photon-induced ones and the colliding objects likely break up. In the case of nucleus emitter, one cannot consider
the entire nucleus charge $Z$ if $b < R$. 

Integrating \ce{eq:dndkd2b} over $b$ down to $b_{\rm min}$, one has~\cite{Baltz:2007kq}
\eqs{
\frac{dn}{dk} = \frac{2 Z^2\alpha_{\rm em}}{\pi k}
\Bigg[
\omega(b_{\rm min},k)K_0\big(\omega(b_{\rm min},k)\big)K_1\big(\omega(b_{\rm min},k)\big) \\
-\frac{{\omega(b_{\rm min},k)}^2}{2}
\big(K_1^2\big(\omega(b_{\rm min},k)\big)-K_0^2\big(\omega(b_{\rm min},k)\big)\Big)
\Bigg].
}

For \pp\ collisions, we choose $b_{\rm min}\simeq 2\times R_p$; 
for \pA\ collisions, $b_{\rm min}\simeq R_p+R_A$; and for $AB$ collisions  $b_{\rm min}\simeq R_A+R_B$
for the number presented in \ct{tab:UPC-parameters}.
Whereas one can approximate $R_p+R_A$ to $R_A$, we do not find appropriate 
to use $R_{\rm Pb}$ for PbPb collisions, for instance. In addition, in \pA\ collisions, it is also 
probably not justified to use a different $b_{\rm min}$ when one considers the proton emission
or the ion emission. In both cases, $R_p+R_A$, or perhaps $R_A$, are to be considered.

\begin{table}[htb!]
\begin{center}\renewcommand{\arraystretch}{1.2}

\scriptsize
\begin{tabular}{l|c|c|c|c|c} 
\hline\hline 
System   & $\sqrt{s_{{NN}}}$ & ${\cal L}_{AB}$\protect\footnotemark 
   & $E_{\gamma~\rm max}^{\rm B\ rest}$  & $\sqrt{s_{\gamma_N}^{\rm max}}$  & $E_{\gamma~\rm max}^{\rm cms}$   \\
 & (GeV) & (pb$^{-1}$yr$^{-1}$) & (GeV) & (GeV)  & (GeV)  \\ \hline
AFTER@LHC\\
$p$H$^\uparrow$   	& 115 & $1.0 \times 10^4$    & 1050  & {44}   & {8.6}   \\ 
$p$D$^\uparrow$   	& 115 & $1.1 \times 10^4$   &  {520}   & {30}   & {4.2}   \\  
$p^3$He$^\uparrow$  & 115 & $3.7 \times 10^4$  	 & {520}   & {30}   & 4.2  \\  
PbH$^\uparrow$ 	    & 72  & 0.12 & {74}  & {12}   & {0.97}  \\ 
PbD$^\uparrow$ 	    & 72  & $8.8 \times 10^{-2}$   & {62}  & {11}   & {0.82}  \\
Pb$^3$He$^\uparrow$ & 72 & $8.3 \times 10^{-2}$   & {62}  & {11}   & {0.82}  \\ \hline

RHIC (STAR)\\
$p^\uparrow p^\uparrow$ (2017)  				& 510 & 400                  & {3190}  & {77}   & {15}     \\ 
Au$^\uparrow p^\uparrow$ (2023)   & 200 & 1.75 & 570 & 33 & 2.7 \\
\end{tabular}
\caption[]{Relevant parameters for  $AB$ UPCs at  AFTER@LHC and RHIC:
(i) nucleon-nucleon cms, $\sqrtsnn$ (ii)  luminosity, $\Lumi_{AB}$, 
(iii) photon ``cutoff energy'' in the target rest frame, $E_{\gamma~\rm max}^{\rm B\ rest}$
(iv) ``maximum'' photon-nucleon cms energy where the $A$ the photon emitter, $\sqrt{s_{\gamma_N}^{\rm max}}$ 
(v) photon ``cutoff energy'' in the cms, $E_{\gamma~\rm max}^{\rm cms}$, with both $A$ and $B$ emitting a photon coherently. Note that we assumed $r_d \simeq r_{^3{\rm He}}$.
}
\label{tab:UPC-parameters}
\end{center}
\end{table}

\footnotetext{For the AFTER@LHC case, the integrated luminosities given are maximum and do not account for possible data taking limitations from the detector point of view. Assuming an LHCb-like detector, the luminosities for  $p$D$^\uparrow$  and  $p^3$He$^\uparrow$ have to be limited to 1.0 $\times$ 10$^{4}$ pb$^{-1}$yr$^{-1}$ and 0.6 $\times$ 10$^{4}$ pb$^{-1}$yr$^{-1}$ respectively.}

From the numbers of the fifth column, it is clear that such photon-nucleon collision are energetic
enough to produce particles like $J/\psi$ as we discuss now.

\begin{table*}[hpbt!]
\footnotesize
\begin{center}\renewcommand{\arraystretch}{1.5}
\begin{tabular}{c|c|c|c|c|c}
\hline \hline
                & Case 1  & Case 2 & Case 3 & \multicolumn{2}{c}{Case 4}   \\ \hline 
 Photon-emitter  &  proton     & lead      &    proton   & proton & gold\\ 
 $\sigma^{\rm tot}_{J/\psi} \times {\rm BR} =\sigma_{J/\psi \rightarrow \ell^{+}\ell^{-}}$ (pb) & $70.10$ & $16.50\times10^{3}$& $364.19$ & $4.76 \times10^{3}$ & $132.50\times10^{3}$ \\ 
$\sigma_{J/\psi \rightarrow \ell^{+}\ell^{-}}$ (with $\eta^\ell$ cut) (pb) & $20.65$ & $9.81\times10^{3}$ & $103.28$ & $2.88 \times10^{3}$ & $23.06\times10^{3}$   \\ 
$\sigma_{J/\psi \rightarrow \ell^{+}\ell^{-}}$ (with $\eta^\ell$ cut, with $P_T$ cut) (pb) & $20.64$ & $9.81\times10^{3}$ & $103.28$ & $2.88\times10^{3}$ & $23.06\times10^{3}$ \\ \hline\hline
\end{tabular}\vspace*{-0.5cm} 
\end{center}
\caption{Summary table of $J/\psi$ photo-production cross sections obtained with the {\sc Starlight} MC generator for four type of collisions (see text for details).}\vspace*{-0.5cm} 
\label{tab_starlight}
\end{table*}

\section{Cross-section and yield estimations with {\sc Starlight}}

\begin{figure*}
\centering
 \subfloat[Case 1]{\includegraphics[width=0.35\linewidth]{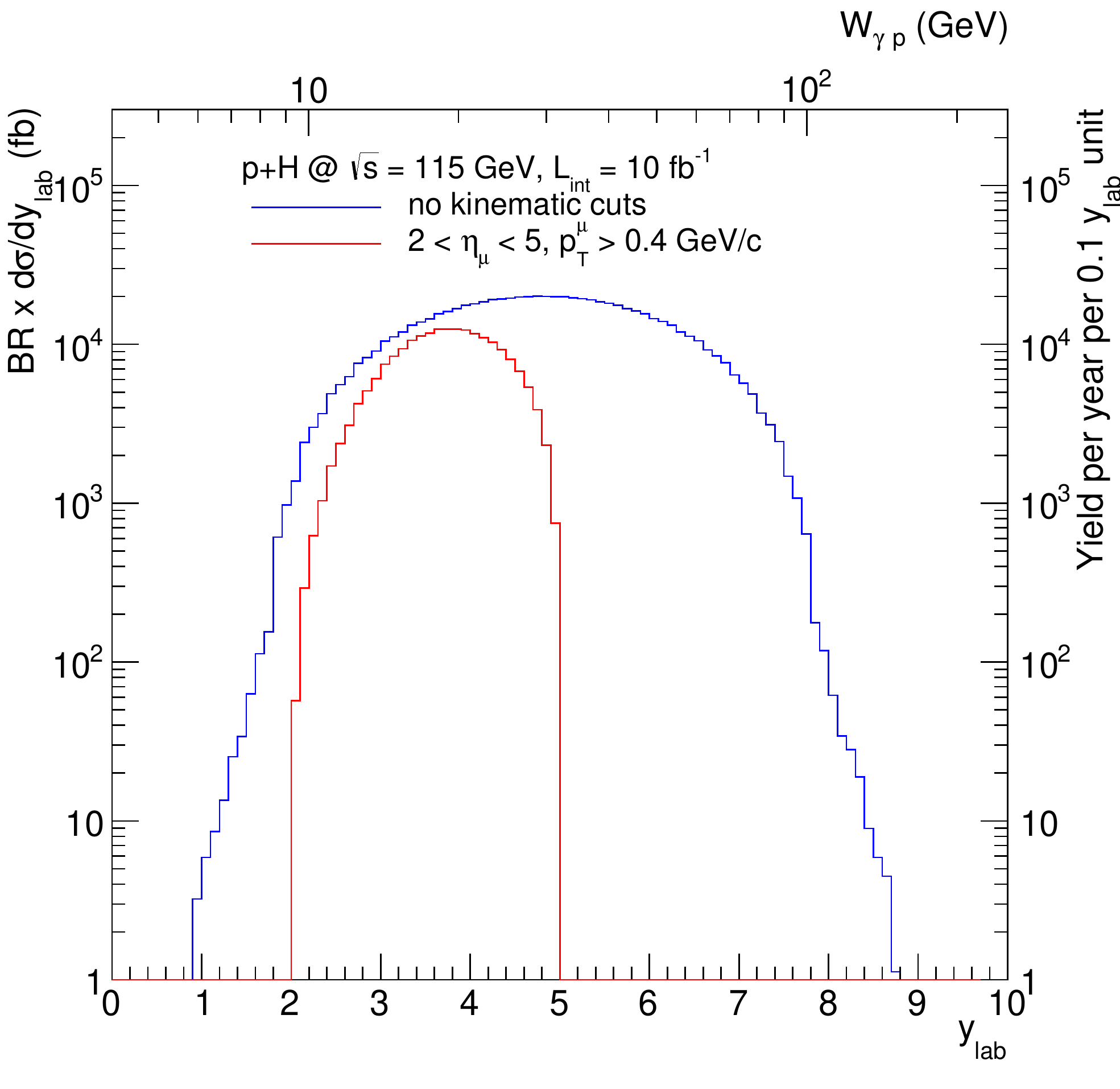}}\quad 
 \subfloat[Case 1]{\includegraphics[width=0.35\linewidth]{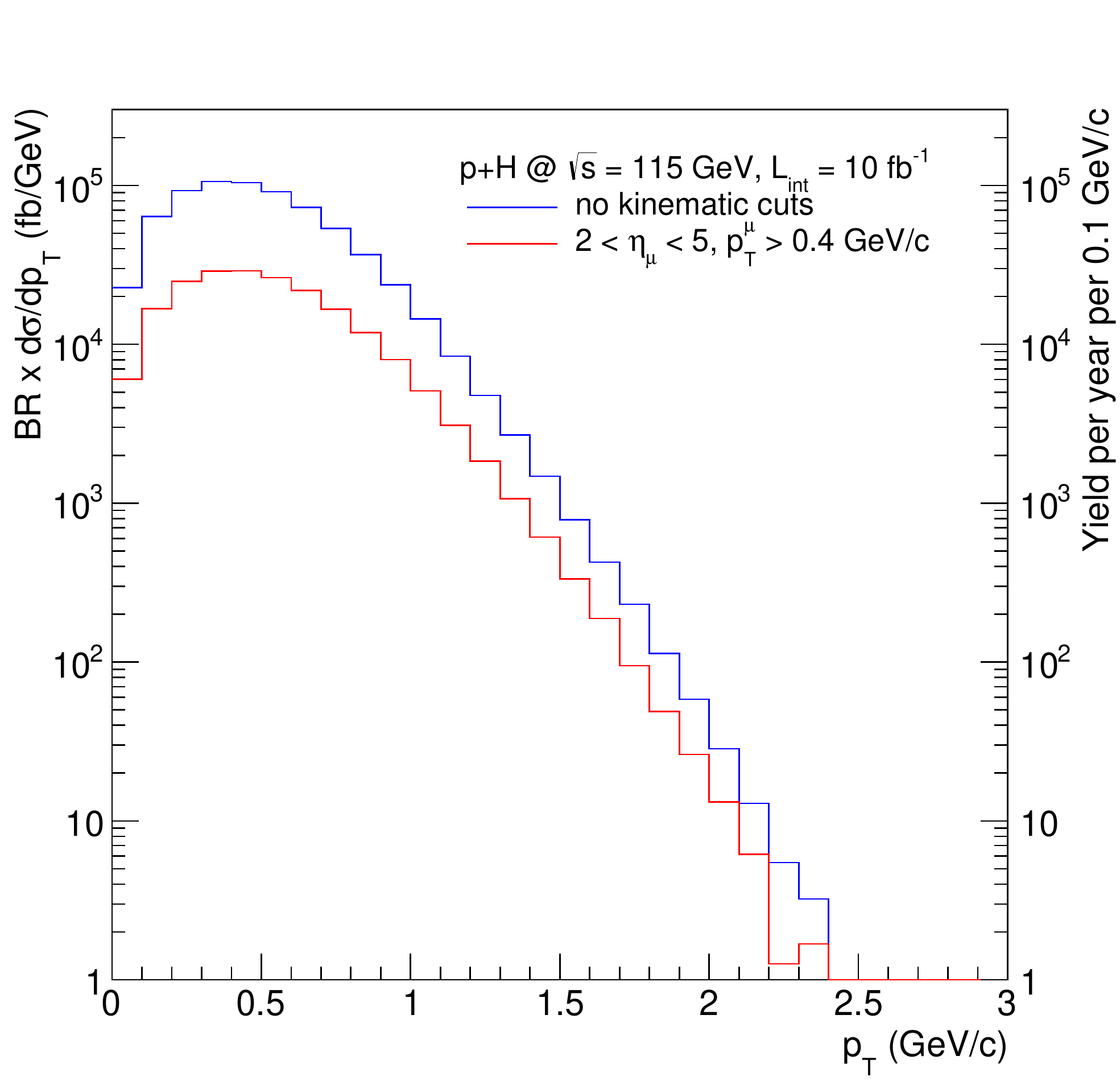}}\vspace*{-0.45cm} \\
 \subfloat[Case 2]{\includegraphics[width=0.35\linewidth]{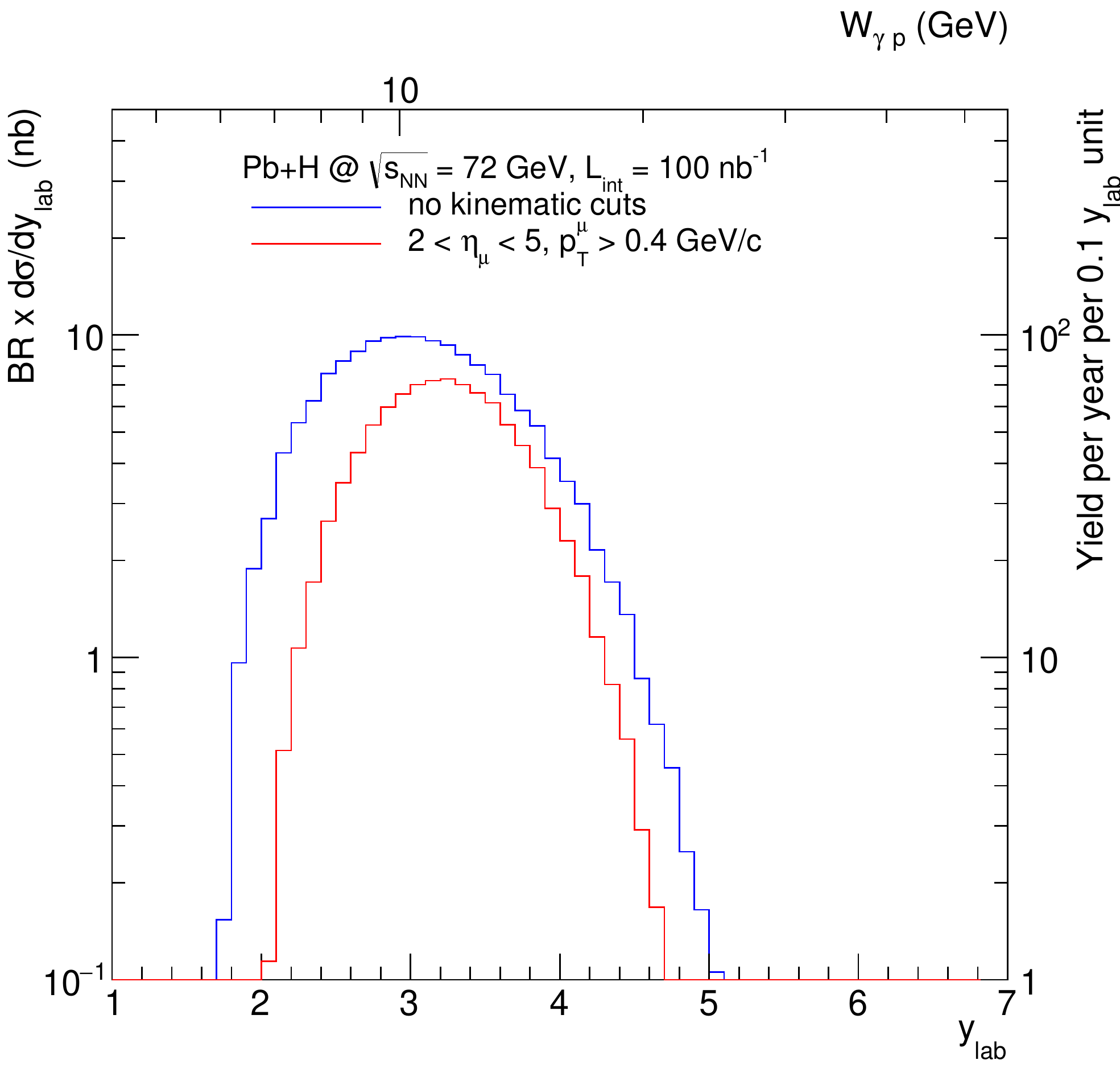}}\quad 
 \subfloat[Case 2]{\includegraphics[width=0.35\linewidth]{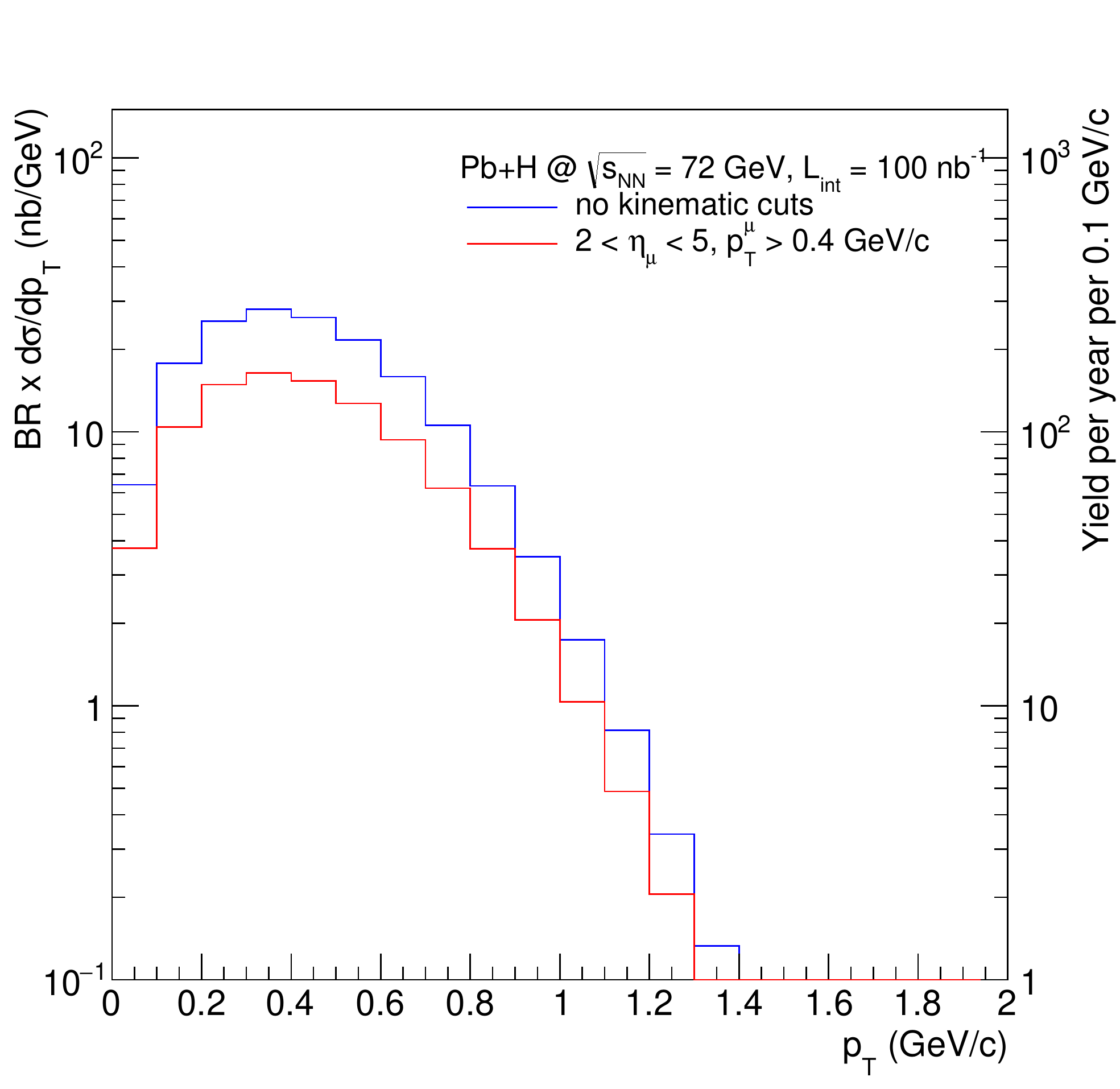}}\vspace*{-0.25cm} 
\caption{$y_{\rm lab}$- (a \& c) and $P_T$-differential (b \& d) $J/\psi$ photo-production cross sections from {\sc Starlight}, for case 1 and 2. The yearly yields are given by the right vertical axis. The blue curves have been produced without applying kinematical cuts, while the red curves are produced by applying the $\eta$ and $P_T$ cuts described in the text. The $W_{\gamma p}$ range probed is also shown on the top axis of the left plot using $W^2_{\gamma p} \equiv {M_\psi^2 + M_N^2 + 2 M_N M_\psi \cosh(y_{\rm lab})}$.}\vspace*{-0.25cm} 
\label{pH_AFTER}       
\end{figure*}

\begin{figure*}[htpb]
\centering
\includegraphics[width=0.5\textwidth,clip]{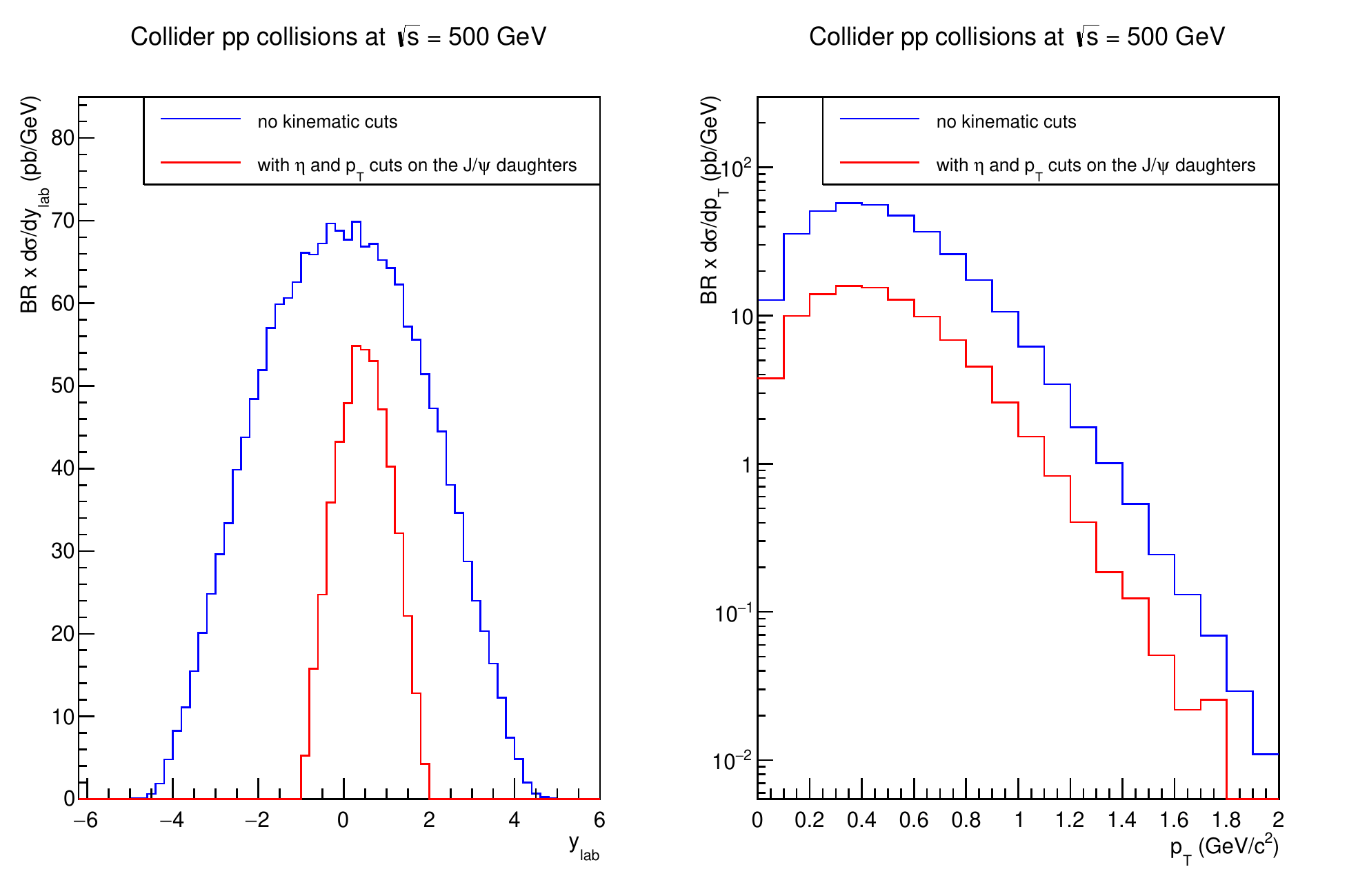}
\caption{$y_{\rm lab}$- (left) and $P_T$-differential (right)  $J/\psi$ photo-production cross sections from {\sc Starlight} for Case 3. The blue curves have been produced without applying kinematical cuts, while the red curves are produced by applying the $\eta$ and $P_T$ cuts described in the text.}
\label{pp_RHIC}       
\end{figure*}

In order to assess the possibility to measure STSAs of exclusively photo-produced $J/\psi$ , we have evaluated the expected rates
with the luminosities and kinematical conditions reported in the previous section using the {\sc Starlight} MC generator
~\cite{Klein:2016yzr} for four type of collisions: 
\begin{enumerate}
\item{$p$H$^\uparrow$ collisions in the fixed-target mode ($\sqrt{s_{NN}}$~=~115~GeV) for AFTER@LHC;}
\item{PbH$^\uparrow$ collisions in the fixed-target mode ($\sqrt{s_{NN}}$~=~72~GeV) for AFTER@LHC; }
\item{$p^\uparrow p^\uparrow$ collisions in the collider mode ($\sqrt{s_{NN}}$~=~500~GeV) for RHIC;}
\item{Au$^\uparrow p^\uparrow$ collisions in the collider mode ($\sqrt{s_{NN}}$~=~200~GeV) for RHIC.}
\end{enumerate}

The first particle is always defined with a positive rapidity (both in fixed-target or collider modes).  For instance, this means that, in Au$p^\uparrow$ collisions, the gold ion travels with a positive rapidity and proton with a negative rapidity (case 4). 
A summary of the production cross sections obtained in the four cases is reported in \ct{tab_starlight}. The second line indicates which particle is considered to be the photon emitter for the cross-section computation. 
The third line reports the production cross section for the dimuons (case 1 and 2) and dielectrons (case 3 and 4) resulting from the $J/\psi$ decay (assuming  a coherent photo-production when the photon-receptor is a nuclei). 
On the fourth line, we reported the same production cross sections after the application of pseudo-rapidity cuts on the  $J/\psi$ decay products (2 $< \eta_{\rm lab}^{\mu^{\pm}} <$ 5 for the AFTER@LHC cases and -1 $< \eta_{\rm lab}^{e^{\pm}} <$ 2 for the RHIC cases). The fifth line still indicates the cross section but with an additional $P_{T}$ cut on both leptons, namely $P_{T} (e^{\pm},\mu^{\pm}) >$ 0.4 GeV/c. Let us note that the effect of the $P_{T}$ cut, after the pseudo-rapidity cut, is negligible. The kinematical selections listed above for the AFTER@LHC cases are meant to mimic an LHCb-like detector set-up~\cite{Hadjidakis:2018ifr} (alos referred to as AFTER@LHCb), while the kinematic selections for the RHIC cases are the ones described in \cite{Aschenauer:2016our} and corresponds to the STAR detector. \\

\cf{pH_AFTER} (a) shows the rapidity\footnote{in the laboratory frame}-differential cross section of the photo-produced $J/\psi$, in the dimuon decay channel, in proton-Hydrogen fixed-target collisions at $ \sqrt{s}$ = 115~GeV (case 1), obtained with the {\sc Starlight} generator.  \cf{pH_AFTER}~(b) shows the $P_T$-differential cross section of the photo-produced $J/\psi$ for case 1. The blue curves have been produced without applying kinematic cuts (similarly to third line of \ct{tab_starlight}), while the red curves are produced by applying the $\eta$ and $P_T$ cuts described in the text above (similarly to last line of \ct{tab_starlight}). The $y_{\rm lab}$- (left) and $P_T$-differential (right)  cross section distributions of photo-produced $J/\psi$ for case 2 (assuming Pb nuclei as photon-emitter), case 3, case 4 (assuming the proton as photon-emitter), case 4 (assuming the Au nuclei as photon-emitter) are respectively shown on \cf{pH_AFTER} (c \& d),  \cf{pp_RHIC} (left \& right), \cf{pAu_RHIC} (top left \& right) and \cf{pAu_RHIC} (bottom left \& right). Moreover, \cf{pp_RHIC_comp} shows the rapidity-differential cross sections of the photo-produced $J/\psi$ for case 4, where the contributions after kinematical cuts from the gold emitter (solid line) and proton emitter (dashed line) are overlaid for comparison. The $J/\psi$ rapidity distribution for both contributions exhibits similar trend as in Figure 2-20 of Reference \cite{Aschenauer:2016our} obtained with the SARTRE MC generator\cite{Toll:2012mb}.

\begin{figure*}[htb]
\centering
\includegraphics[width=0.5\textwidth,clip]{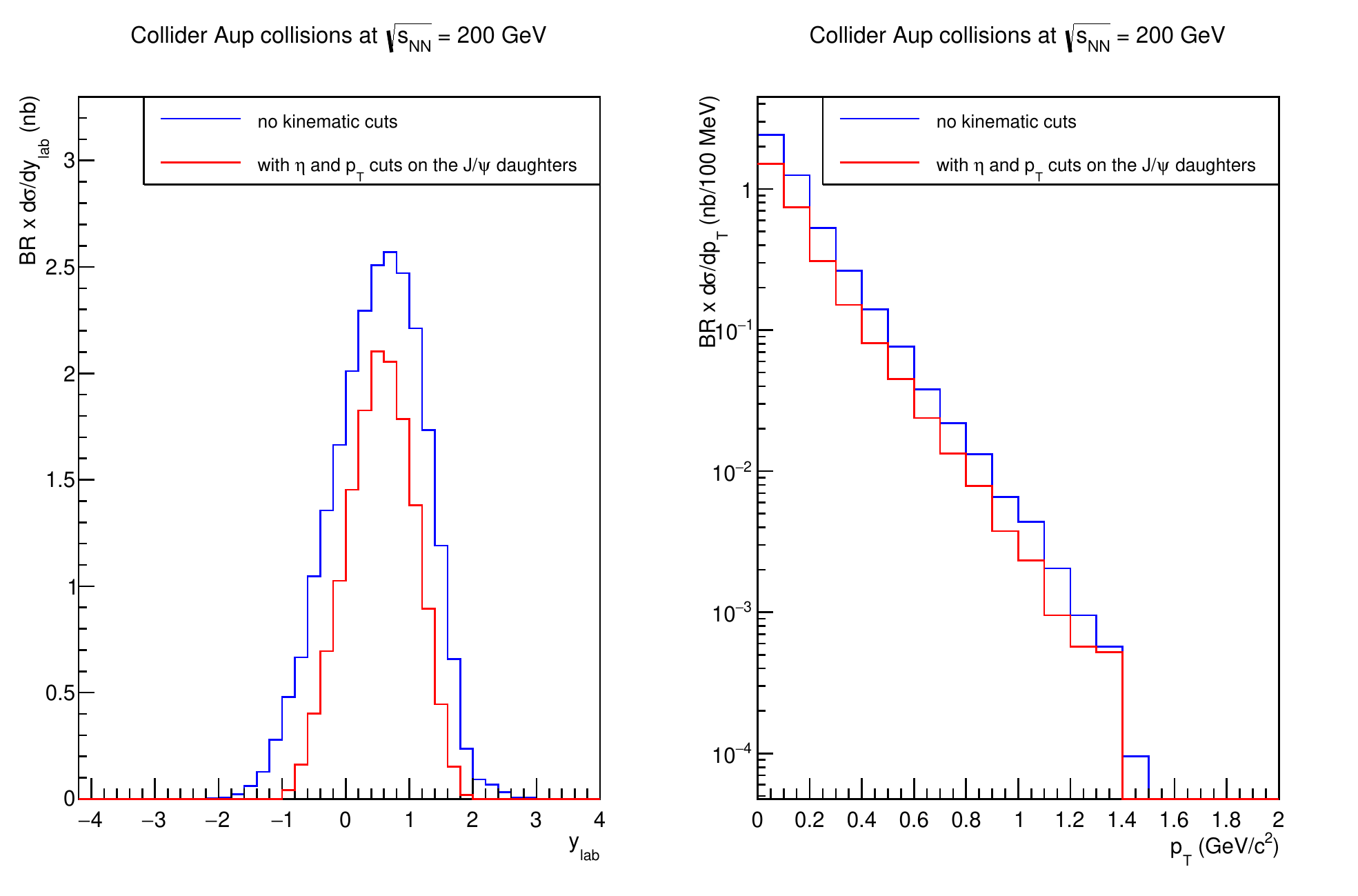}
\includegraphics[width=0.5\textwidth,clip]{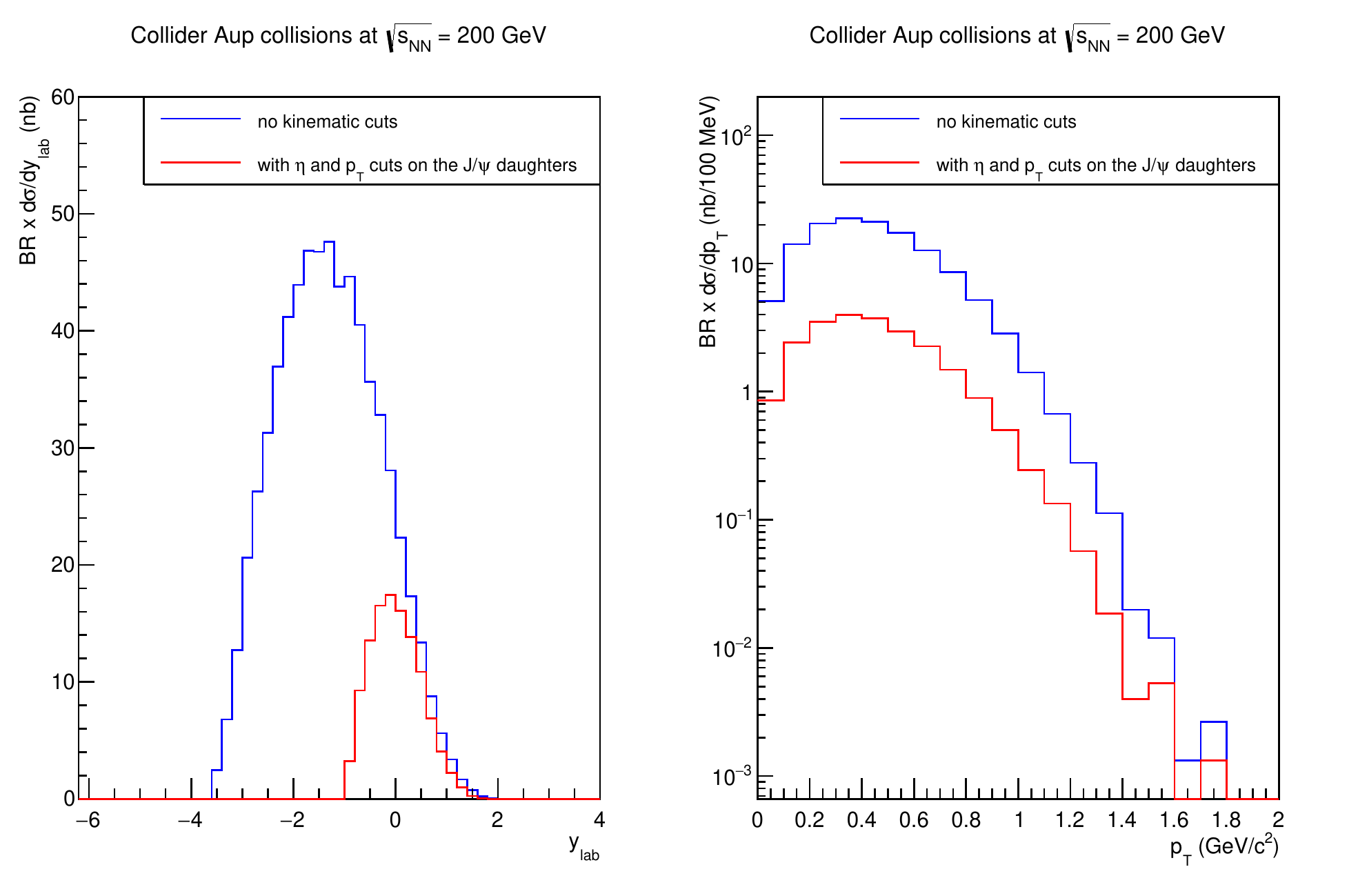}
\caption{$y_{\rm lab}$- (left) and $P_T$-differential (right)  $J/\psi$ photo-production cross sections from {\sc Starlight} for Case 4 (RHIC).
The upper plots correspond photon-emission from the proton and the lower plots from the gold nuclei. The blue curves have been produced without applying kinematical cuts, while the red curves are produced by applying the $\eta$ and $P_T$ cuts described in the text.}
\label{pAu_RHIC}       
\end{figure*}

\begin{figure}[!htb]
\centering
\includegraphics[width=0.35\textwidth,clip]{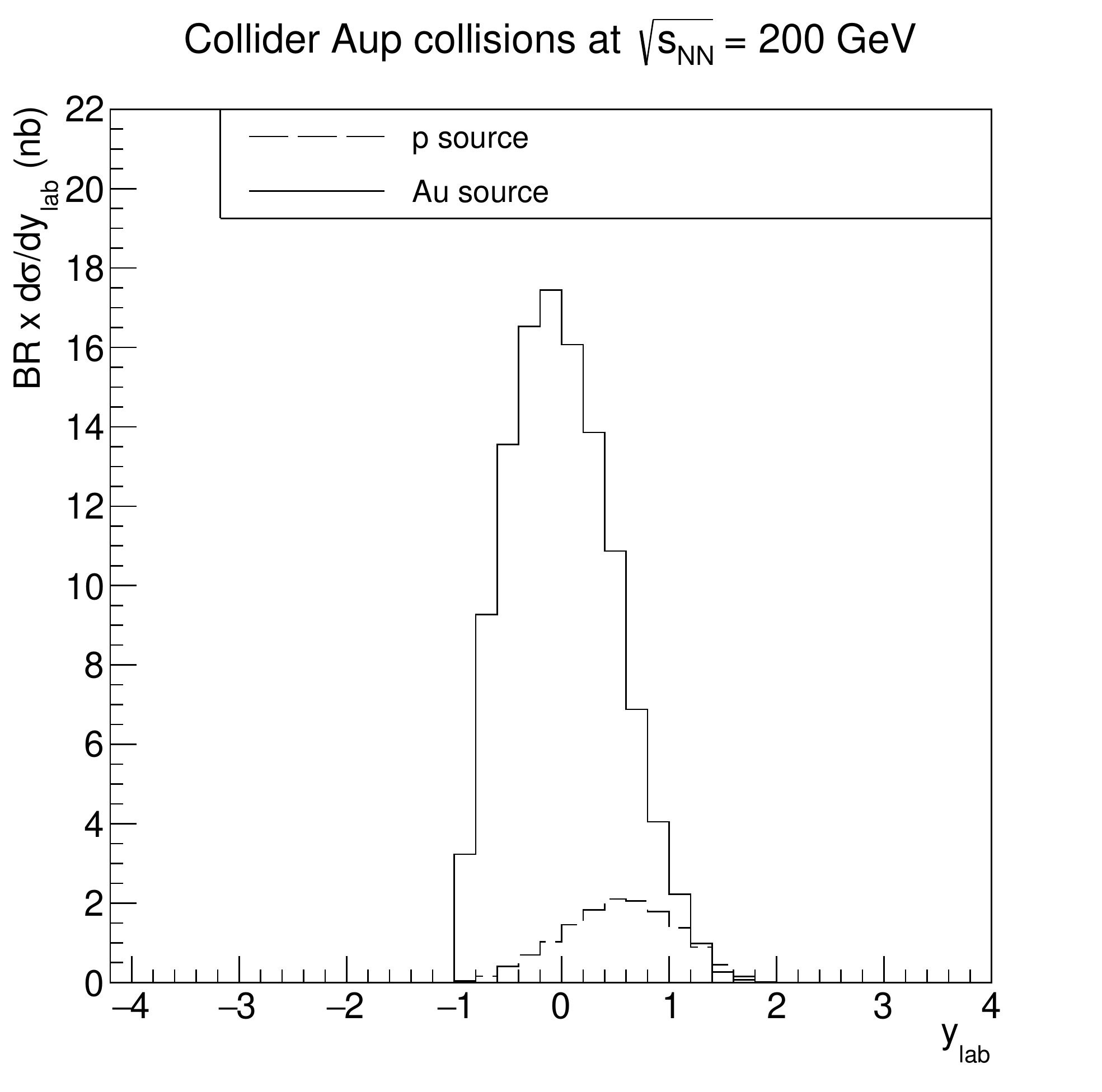}
\caption{Rapidity-differential cross sections of the photo-produced $J/\psi$ in the laboratory frame, from {\sc Starlight} generator for case 4. The dashed line corresponds to the case where the proton is the photon-emitter, the solid line corresponds to the case where the gold nuclei is the photon-emitter. Kinematical selections described in the text were applied to both contributions.}
\label{pp_RHIC_comp}      
\end{figure}

Assuming a polarised internal gas target for AFTER@LHC, with a storage cell like the HERMES system, integrated luminosities as large as 10 fb$^{-1}$ per year would be collected in proton-hydrogen collisions~\cite{Barschel:2015mka,Lansberg:2016urh,Hadjidakis:2018ifr,Redaelli:2018oqa}. This would result in a yearly yield of $\sim$ 200 000 photo-produced $J/\psi$ emitted in the LHCb acceptance. Concerning collisions of Pb nuclei on hydrogen target, the collection of an integrated luminosity of 0.1 pb$^{-1}$ per year is expected in AFTER@LHC with the internal gas target option. This would result in $\sim$ 1 000  photo-produced $J/\psi$ per year\footnote{An LHC year corresponds to about $\sim$ 10$^{6}$s of Pb beam and $\sim$ 10$^{7}$s of proton beam.} emitted in the LHCb acceptance. Since a gas target without a storage cell --like the H-jet system used at RHIC~\cite{Zelenski:2005mz}-- corresponds to luminosities close to 2 order of magnitude lower, it seems difficult (despite a better gas polarisation) to envision such a solution for the PbH$^\uparrow$ case since the flux of polarised hydrogen is limited. Note however that for polarised $^3$He$^\uparrow$ or unpolarised hydrogen, the flux can be increased to compensate for the decrease of luminosity~\cite{Barschel:2015mka,Hadjidakis:2018ifr}.

These numbers can be compared to the expected photo-produced $J/\psi$ yields from simulations, applying the STAR experiment at RHIC kinematical cuts, for $pp$ collisions at $\sqrt{s}$ = 500 GeV. we assumed the Run-2017 STAR data taking conditions \footnote{Note however the slight difference in the  $\sqrt{s}$ assumed, since the simulations were performed prior to the 2017 data taking}, where the collection of an integrated luminosity of 400 pb$^{-1}$ occured. According to our {\sc Starlight} simulations, one could expect the production of about 41~000 $J/\psi$ in the STAR acceptance\footnote{In~\cite{Aschenauer:2016our}, a similar study was performed with the SARTRE MC generator, accounting for, on top of kinematical cuts, all trigger and reconstruction efficiencies. The expected number of detected photo-produced $J/\psi$ was found to be 11000.}. Our simulations suggest that the $J/\psi$ photo-production rate in $p$H$^\uparrow$ collisions at AFTER@LHC is about a factor five bigger that at RHIC per year.

In 2023, STAR is expected to collect 1.75 pb$^{-1}$ of Au$^\uparrow p^\uparrow$ collisions. According to {\sc Starlight}, one would expect the production of  40 000  $J/\psi$\footnote{In~\cite{Aschenauer:2016our}, a similar study was performed with the SARTRE MC generator, accounting for, on top of kinematical cuts, all trigger and reconstruction efficiencies. The expected number of detected photo-produced $J/\psi$ was found to be 13000.} with gold nuclei as the photon source. In PbH$^\uparrow$ collisions, the $J/\psi$ photo-production yield at AFTER@LHC would be smaller by at least one order of magnitude with respect to RHIC  Au$^\uparrow p^\uparrow$ collisions.

\section{Evaluation of the STSAs within the GPD formalism}
The most common theoretical framework to describe exclusive photo-production of vector quarkonia~\cite{Ivanov:2004vd} in the collinear factorisation is based on the introduction of generalised parton distributions (GPDs)~\cite{Guidal:2013rya,Boffi:2007yc,Belitsky:2005qn,Ji:2004gf,Diehl:2003ny,Mueller:1998fv, Ji:1996ek,Radyushkin:1997ki}. 
In ths section, we derive the relation between the $J/\psi$ STSA and the gluon GPDs.

\subsection{Elements of kinematics}

According to \cf{fig:gammap_to_psip}, $q$ is the photon momentum, 
$p$ (resp. $p'$) is the incoming (resp. outgoing) proton momentum
and $p_\psi$ the $J/\psi$ momentum. 
Then, we define
\begin{align}
&
\Delta=p^\prime -p \, , \ \ P=\frac{p+p^\prime}{2} \, , W_{\gamma p}=\sqrt{s_{\gamma p}},  \ t=\Delta^2 \, ,
\nonumber \\
&
(q-\Delta )^2=p_\psi^2=M_\psi^2 \, , \ \ \xi =\frac{M_\psi^2}{2W_{\gamma p}^2 -M_\psi^2}, 
\label{not1}
\end{align}
where $\xi$ is the fraction of the longitudinal momentum transfer. 
\begin{figure}[hbt!]
\centering
\includegraphics[width=0.8\columnwidth,clip]{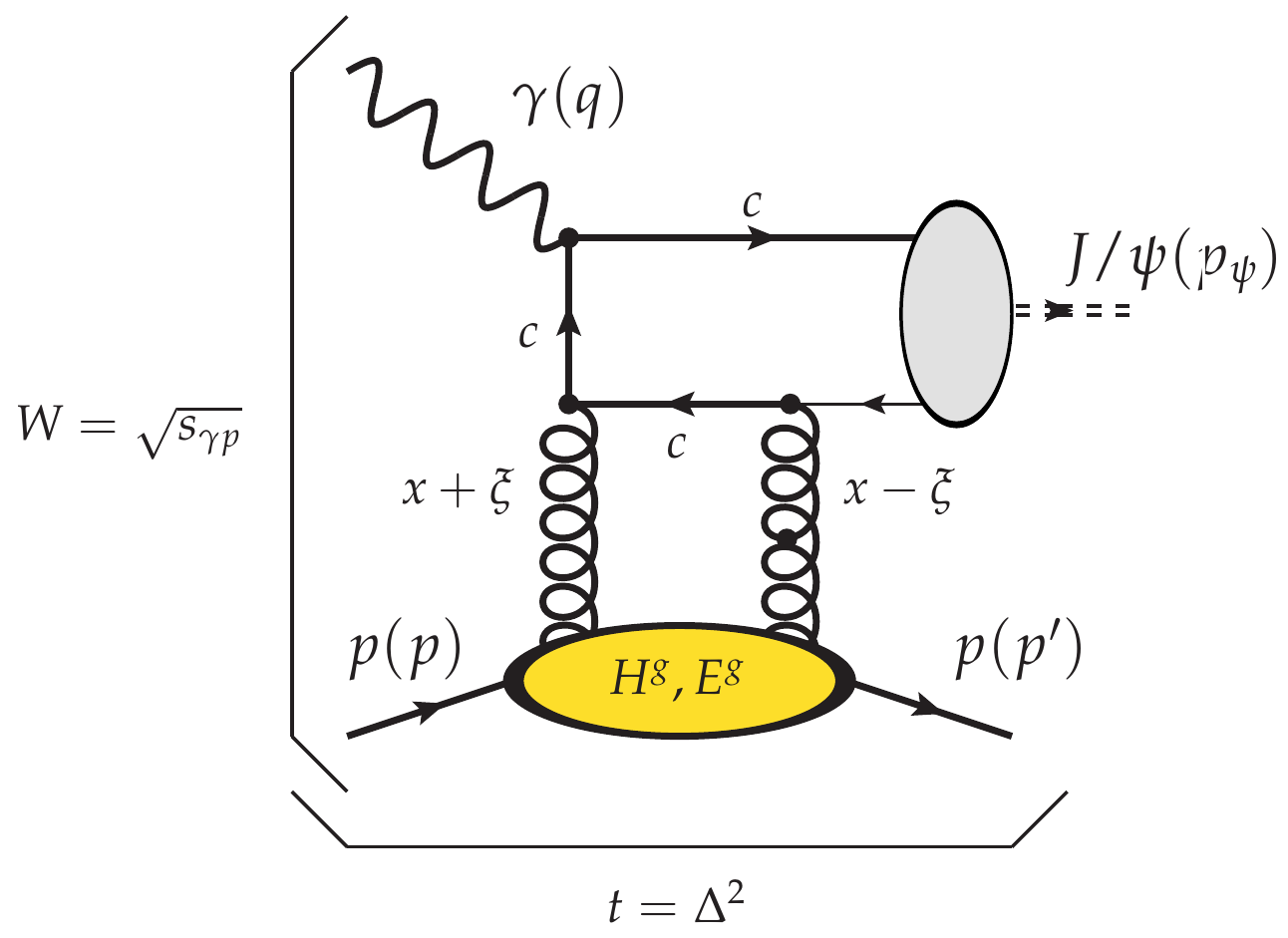}
\caption{Typical Feynman graph for the Born (LO) contribution 
to $J/\psi$ photo-production off a proton with a gluon GPD.\label{fig:gammap_to_psip}	}
\end{figure}

To parametrise the momenta of the particles in the process, 
it is convenient to introduce two light-cone vectors:
$
n_{+}^2=n_{-}^2=0 \, , \ \ n_+ n_- = 1 \, .
$
Any vector $a$ is then decomposed in the following way:
$
a^\mu=a^+n_+^\mu+a^-n_-^\mu+a_\perp \, , \ \ a^2=2a^+a^- - \vectc a \, .
$

We choose the coordinate frame in which the momenta are given by:
\begin{align}
&q=\frac{(W_{\gamma p}^2-m_N^2)}{2(1+\xi)W_{\gamma p}}\, n_- \, ,
p=(1+\xi)W_{\gamma p}\, n_+ + \frac{m_N^2}{2(1+\xi)W_{\gamma p}}\, n_- \, ,
\nonumber \\
&p^\prime=(1-\xi)W_{\gamma p}\, n_+ 
+\frac{(m_N^2+\vectc \Delta)}{2(1-\xi)W_{\gamma p}}\, n_- +\Delta_\perp \, ,
\nonumber \\
&\Delta=-2\, \xi \, W_{\gamma p}\, n_+ +\left(\frac{\xi\, m_N^2}{(1-\xi^2)W_{\gamma p}}+\frac{\vectc
\Delta}{2\,(1-\xi)W_{\gamma p}}
\right)n_- +\Delta_\perp \, ,
\label{not33}
\end{align}
where $m_N$ is the nucleon mass. We are interested in the kinematic region where
the invariant transferred momentum, 
\begin{equation}
t=\Delta^2=-\left(\frac{4\, \xi^2}{1-\xi^2}m_N^2+\frac{1+\xi 
}{1-\xi}\vectc \Delta \right) \, ,
\label{not4}
\end{equation} 
is much smaller (in absolute value) than $M^2_\psi$. In the scaling limit the variable $\xi$ parametrises the 
plus component of the momentum transfer.

\subsection{The STSA in terms  of the GPDs}

The factorisation formula at the leading order in $\alpha_s$, in which the quark contribution is absent, reads:
\begin{align}
&
 {\cal M}
=\frac{2^3 \sqrt{\pi} e e_q \ep^\star(p_\psi)\cdot\ep(p_\gamma)}
{M_\psi^{3/2}\sqrt{N_c} \, \xi} {R(0)}
\int\limits^1_{-1} dx \,
 T_g( x,\xi)\, F^g(x,\xi,t)
\label{fact1}
\end{align}
where:
 $e$ is the electric charge of the heavy quark ($e_c=2/3$,
$e_b=-1/3$), $R(0)$ is the $J/\psi$ radial wave function at the origin in the configuration space, 
$\ep(p_\psi)$ (resp. $\ep(p_\gamma)$) is the polarisation vector of the $J/\psi$ (resp. $\gamma$)
and  $T_g(x,\xi)$ the gluon hard-scattering amplitude,
 describing the partonic subprocesses $\gamma g\to (\bar c c) g$ which, at LO, reads:
\begin{align}
&
 T_g(x,\xi)=\frac{\xi}{(x-\xi+i\varepsilon)(x+\xi-i\varepsilon)}
\alpha_s(\mu_R)\, .
\label{gAT}
\end{align}
The hard-scattering amplitudes 
at NLO were calculated in \cite{Ivanov:2004vd,Jones:2015nna,Jones:2016ldq}.

The relevant GPDs are defined as the matrix element of renormalised light-cone gluon operators and are given by:
\begin{align}
&F^g (x,\xi,t,\mu_F) \label{gGPD}
\\&=\frac{1}{(Pn_-)}\int\frac{d\lambda}{2\pi}
e^{i x (P z)}\,
n_{-\alpha}n_{-\beta}\,\Big\langle
p^\prime \Big|G^{\alpha\mu} \left(-\frac{z}{2}\right)
G^{\beta}_\mu \left(\frac{z}{2}\right)
\Big|p\Big\rangle\Big|_{z=\lambda n_-}
\nonumber \\
&
=\frac{1}{2(Pn_- )}\Big[
H^g \, \bar u(p^\prime)\not \! n_- u(p)+
E^g \, \bar
u(p^\prime)\frac{i\sigma^{\alpha\beta}n_{-\alpha}\Delta_\beta}{2\,m_N} 
u(p)
\Big],\nn
\end{align}
where $H^g$ and $E^g$ are functions of $x$, $\xi$, $t$ and of
the factorisation scale $\mu_F$. In the current study, owing the lack of knowledge on the GPD $E^g$, we do not consider useful to study the scale uncertainties by varying them about a default value. A reasonable value for the latter is $M_{\psi}$ which we use for $\mu_R$ and $\mu_F$. This choice is implied in the following formulae. We further note that the insertion of a path-ordered gauge factor between the
field operators is implied in the above definition. We do not discuss it further as it does not affect the phenomenology.

To go further we introduce the gluonic form factors
\begin{align}
{\cal H}^g(\xi,t) \equiv \int\limits_{-1}^1dx\; T_g(x,\xi)H^g(x,\xi,t),\nn\\
{\cal E}^g(\xi,t) \equiv \int\limits_{-1}^1dx\; T_g(x,\xi)E^g(x,\xi,t).
\label{ffactors}
\end{align}
which permit to write gluonic contribution 
to the scattering amplitude as\footnote{In what follows, 
we will drop $\xi,t$ dependence of the gluonic form factors.}
\begin{align}
 {\cal M}
&=\frac{2^3 \sqrt{\pi} e e_q \ep^\star(p_\psi)\cdot\ep(p_\gamma)}
{M_\psi^{3/2}\sqrt{N_c} \, \xi}
\cdot \frac{R(0)}{2(Pn_- )}
\nonumber \\&\left[
 {\cal H}^g \bar u(p^\prime)\!\not \! n_- u(p)+
{\cal E}^g  \bar
u(p^\prime)\frac{i\sigma^{n_{-}\Delta}}{2\,m_N} 
u(p)
\right]\;.
\label{amplffactors}
\end{align} 
In order to calculate transverse spin asymmetry, we assume that initial proton is polarised and characterised by the polarisation vector $s^\mu$ such that $p\cdot s=0$ and $s^2=-1$, then using
$u(p,s)\bar u(p,s) = \frac{1}{2}(\hat p + m_N )(1+\gamma^5 \!\not \! s)$
the $s$ dependent part [hence the subscript "$s$"] of the square of absolute value of spinor matrix element in \ce{amplffactors} (in bracket) summed over the polarisation of the final nucleon reduces to
\begin{align}
&\sum_{s'}\Bigg| \Bigg[
 {\cal H}^g\bar u(p^\prime,s')\!\not \! n_- u(p,s)+
{\cal E}^g  \bar
u(p^\prime,s')\frac{i\sigma^{n_{-}\Delta}}{2\,m_N} 
u(p,s)
\Bigg]\Bigg|^2_s
\nonumber \\
&= -\frac{2(Pn_- )}{m_N}(1+\xi) \epsilon^{n_-\,n_+\,s\, \Delta_T}\,\Im[{\cal H}^g{{\cal E}^g}^\star].
\label{polsquare}
\end{align}
where ($\epsilon^{0123}=1$)
\begin{eqnarray}
\epsilon^{n_-\,n_+\,s\, \Delta_T} = -\frac{1}{2}(s_x\Delta_y - s_y\Delta_x)=-\frac{1}{2}|\Delta_T| \sin( \phi_{\vect \Delta} )\;.
\label{epsilon}
\end{eqnarray}
One sees that the spin dependence only survives when the polarisation vector $s$ has a transverse component, \ie\ the initial proton is (transversely) linearly polarised. In what follows, we will assume that the proton is transversely polarised. The $ \phi_{\vect \Delta}$ is the angle between $\vect \Delta$ and $s$ vectors.

On the other hand, the spinor matrix element in bracket in \ce{amplffactors} summed over the polarisations both of initial and final protons equals
\begin{align}
&\!\!\sum_{s'\,s}\Bigg| 
 {\cal H}^g \bar u(p^\prime,s')\!\not \! n_- u(p,s')+\label{unpolar}
{\cal E}^g  \bar
u(p^\prime,s')\frac{i\sigma^{n_{-}\Delta}}{2\,m_N} 
u(p,s)
\Bigg|^2\!\!\!
 \\\nn
 & = 4(Pn_- ) \Big[  (1-\xi^2) |{\cal H}^g|^2
+\frac{\xi^4}{1-\xi^2}|{\cal E}^g|^2 -2 \xi^2 \Re({\cal H}^g{\cal E}^{g\,*}	)  \Big]\;.\nn
\end{align}
The expresions of \ce{polsquare} and \ce{unpolar}) constitute the basis of definition of the STSAs, 
which in the photo-production case, can be written in the form 
\begin{align}
&\!\!{A}^\gamma_N =\frac{\sum\limits_{s,s'}s_x \Big| 
 {\cal H}^g \bar u(p^\prime,s')\!\not \! n_- u(p,s)+
{\cal E}^g  \bar
u(p^\prime,s')\frac{i\sigma^{n_{-}\Delta}}{2\,m_N} 
u(p,s)\Big|^2 
}{\sum\limits_{s'\,s}\Big|
 {\cal H}^g \bar u(p^\prime)\not \! n_- u(p)+
{\cal E}^g  \bar
u(p^\prime)\frac{i\sigma^{n_{-}\Delta}}{2\,m_N} 
u(p)
\Big|^2}
\nonumber \\
& =\frac{\frac{1}{2m_N }(1+\xi) |\Delta_T| \sin( \phi_{\vect \Delta})\,\Im({\cal H}^g{\cal E}^{g\, \star})}{ (1-\xi^2) |{\cal H}^g|^2 +\frac{\xi^4}{1-\xi^2}|{\cal E}^g|^2 -2 \xi^2 \Re({\cal H}^g{\cal E}^{g\,\star})}\;.
\label{stsa}
\end{align}
The GPD ${ H^g}$ is and will be extracted from exclusive processes with unpolarised target. Well tested models describing it exists. On the other hand, almost nothing is known about the GPD $E^g$ which plays crucial role in the Ji sum rule describing decomposition of the proton spin \cite{Ji:1996ek}. Equation \ref{stsa}, previously obtained in \cite{Koempel:2011rc}, proves that measuring STSAs should significantly improve that knowledge. 

\subsection{STSA magnitude prediction and uncertainty projections for AFTER@LHC}

To estimate size of the expected asymmetry we will use the popular Goloskokov-Kroll model for the GPD $H^g$ \cite{Kroll:2012sm}. As what concerns the essentially unknown GPD $E^g$, we will following the modelling of~\cite{Koempel:2011rc}, and choose the variant V4 resulting in the largest asymmetry. This choice should however not be seen as a potential upper limit. Indeed, since this paper appeared, relatively large gluon-Sivers based spin asymmetries were observed by COMPASS in di-hadron production~\cite{Adolph:2017pgv}. As such, $E^g$
cannot be negligibly small as sometimes thought earlier.

\begin{figure}[hbt!]
\includegraphics[width=\columnwidth]{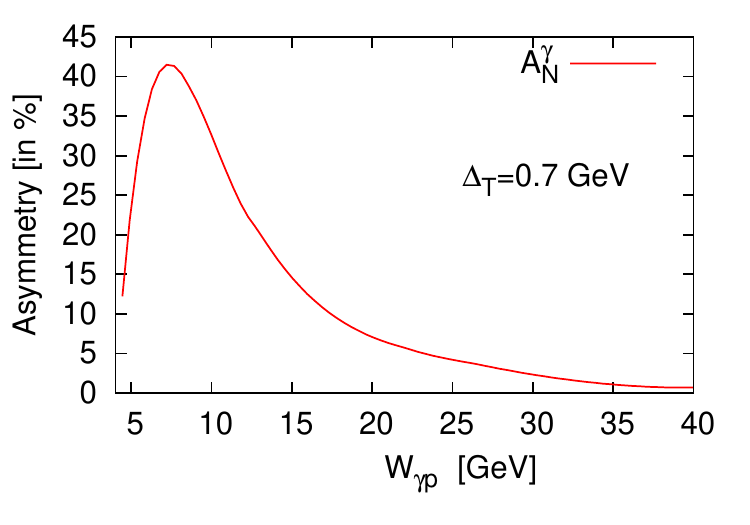}
\caption{STSA in photo-production for $\Delta_T = 0.7\textrm{~GeV}$.}
\label{fig:AsymmPhoto}
\end{figure}  
On \cf{fig:AsymmPhoto}, we show the magnitude of the asymmetry in the photo-production $A^{\gamma}_N$ as a function of $W_{\gamma p}$, for $\Delta_T=0.7 \textrm{~GeV}$. We observe that the used models predict a sizeable asymmetry for moderate values of $W_{\gamma p}$ and its gets close to zero for larger energies.

AFTER@LHC would create a unique possibility to study such a single transverse spin asymmetries, which is sensitive to yet unknown GPD $E_g$\cite{Koempel:2011rc}, through UPCs. In the present analysis we study two modes: proton-hydrogen and lead-hydrogen collisions at AFTER@LHC. Using the Equivalent Photon Approximation (EPA) we can calculate the hadronic cross section as the convolution of the Weizsacker-Williams photon fluxes with the photo-production cross section:
\begin{align}
\sigma^{h_A h_B} = \int dk \left[\frac{dn_A}{dk}\sigma^{\gamma h_B} +\frac{dn_B}{dk}\sigma^{\gamma h_A} \right]
\end{align}
Assuming that the hadron $B$ is polarised, the (hadronic) STSA, $A_N$ can be expressed in terms of the (photonic) STSA $A^\gamma_N$ :
\begin{align}
A_N &= \frac{\sigma^{h_A h_B^\downarrow}-\sigma^{h_A h_B^\uparrow}}{\sigma^{h_A h_B^\downarrow}+\sigma^{h_A h_B^\uparrow}}
=\frac{
\int dk \frac{dn_A}{dk}\left[\sigma^{\gamma h_B^\downarrow}-\sigma^{\gamma h_B^\uparrow}\right]
}{\sigma^{h_A h_B}}\\
&=
\frac{
\int dk \frac{dn_A}{dk}
\frac{
\sigma^{\gamma h_B^\downarrow}-\sigma^{\gamma h_B^\uparrow}
}
{\sigma^{\gamma h_B}
}
 \sigma^{\gamma h_B}
}{\sigma^{h_A h_B}}
=
\frac{
\int dk \frac{dn_A}{dk}
A^\gamma_N~	 \sigma^{\gamma h_B}
}{\int dk \left[\frac{dn_A}{dk}\sigma^{\gamma h_B} +\frac{dn_B}{dk}\sigma^{\gamma h_A} \right]}\nn
\end{align}

To get the most realistic predictions for the asymmetry in UPCs, we are using the  GPD-based prediction for $A^\gamma_N$ given by \ce{stsa}. However, it is well known that the normalisation of the $J/\psi$ production cross section based on GPDs is plagued by large uncertainties. 
Since data exist, it is therefore expedient to rather resort to a parametrisation of the unpolarised cross section like the one used in {\sc Starlight}~\cite{Klein:2016yzr}, namely
\eqs{\sigma(\gamma + p \to J/\psi + p)= \sigma_P \Bigg[1 - \frac{(m_N+m_{J/\psi})^2}{W^2_{\gamma p}} \Bigg]^2 W^\epsilon_{\gamma p} }
with $\sigma_P=4.06\text{ nb}$ and $\epsilon=0.65$.
The $y$ distribution and $p_T$ distribution for those cases are shown on the Fig.\ref{pH_AFTER}.

Our prediction for the STSA along  with its statistical uncertainty\footnote{These are evaluated as follows. 
The photo-produced $J/\psi$ yields, $N$ obtained with {\sc Starlight} and the evaluated magnitude of the STSA, defined as
\begin{equation}
A_{N} = \frac{1}{P_{\rm eff}}\frac{N^{\uparrow} - N^{\downarrow}}{N^{\uparrow} + N^{\downarrow}},
\end{equation}
allows one to evaluate $N^{\uparrow}$ and $N^{\downarrow}$ ($N=N^{\uparrow}+N^{\downarrow}$), \ie\ the number of photo-produced $J/\psi$ for an up (down) target polarisation orientation where $P_{\rm eff}$ is the effective polarisation of the target. From these, we have evaluated the statistical uncertainty on $A_{N}$, $\delta_{A_{N}}$, as  
\begin{equation}
\delta_{A_{N}} = \frac{2}{P_{\rm eff}(N^{\uparrow} + N^{\downarrow})^{2}}\sqrt{N^{\downarrow2}\delta^{\uparrow2} + N^{\uparrow2}\delta^{\downarrow2}}
\end{equation}
with $\delta^\uparrow$ ($\delta^\downarrow$) the relative uncertainties on the $J\psi$ yields with up (down) polarisation orientation.}
in the kinematics relevant for the GPD extraction are presented as a function of Feynman-$x$, $x_F$,\footnote{$x_F$ is defined as: 
$
x_F = 2  (M_{J/\psi}/\sqrt{s}) \sinh ({y_{cms}}),
$
where  $y_{cms}$ is the $J/\psi$ rapidity in the cms frame and $\sqrt{s}$ the cms energy.} on \cf{AN_final} . It clearly indicate that AFTER@LHC is able to perform the first determination of $E_g$.

\begin{figure}[!hbt]
\centering
\subfloat[$p$H$^\uparrow$]{\includegraphics[width=9.0cm,clip]{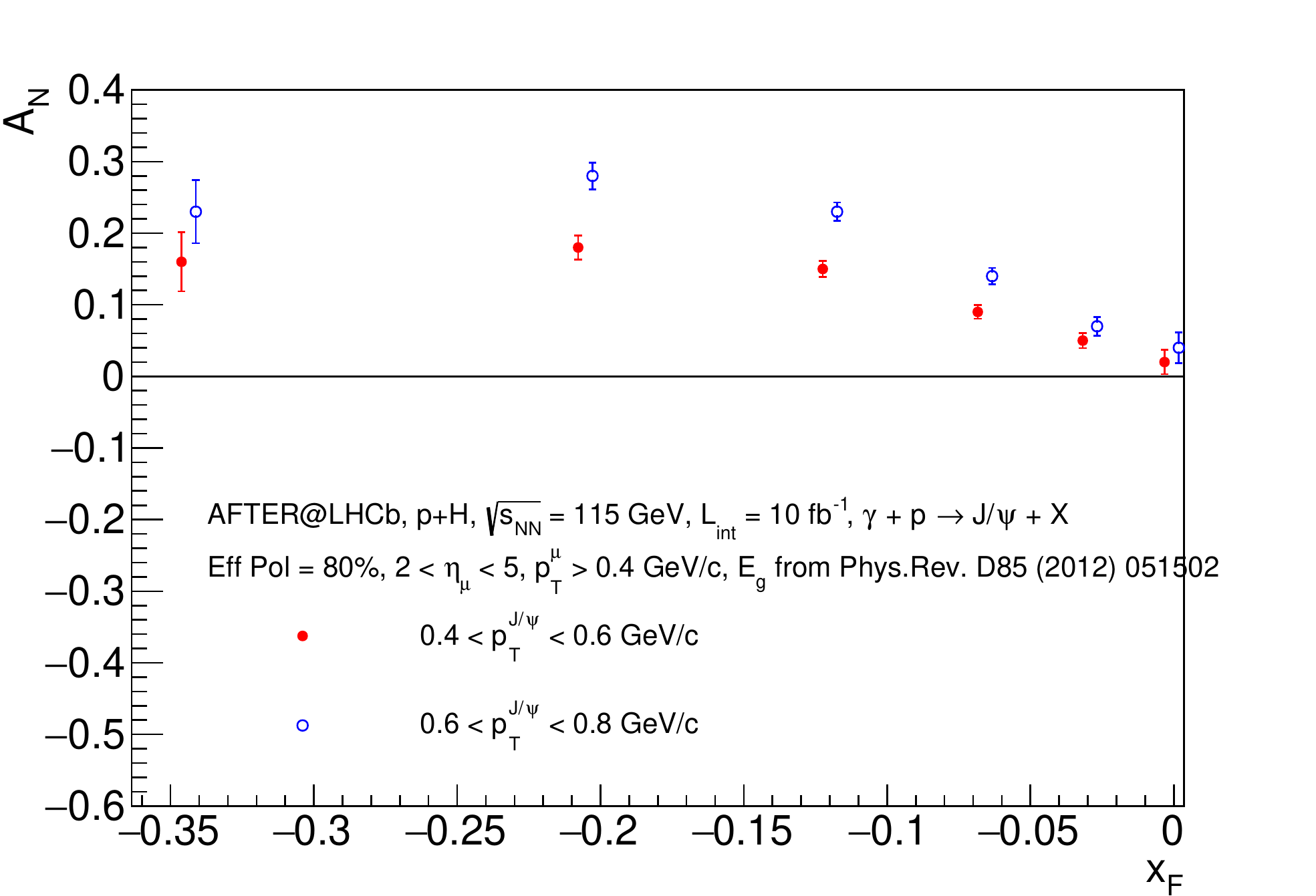}}\\
\subfloat[PbH$^\uparrow$]{\includegraphics[width=9.0cm,clip]{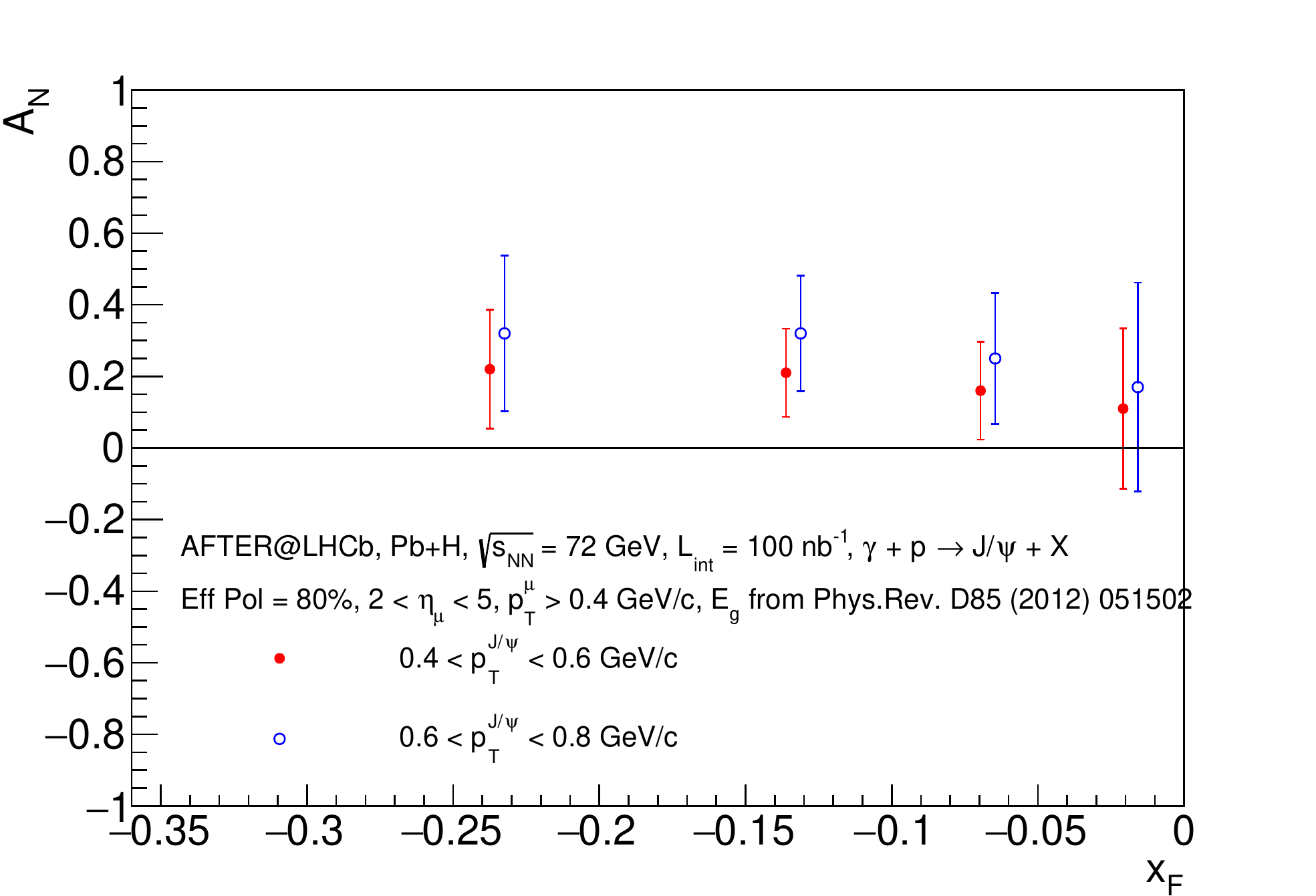}}
\caption{STSAs in the exclusive $J/\psi$ photo-production in UPCs with a proton beam (a) and a lead beam (b) on an transversely polarised hydrogen target.}
\label{AN_final}      
\end{figure}

\section{Conclusions}

In conclusion, we have evaluated the expected cross sections for a LHCb-like detector used in fixed-target mode (AFTER@LHCb)
with the 7 TeV $p$ and 2.76 TeV Pb LHC beams and compared them to those expected at RHIC. These are similar. However, the use of the fixed-target mode allows one to probe a very different kinematics at much larger $x$ in the polarised nucleons.

Using a polarised-internal-gas target with a storage cell, we expect to be able to record a fraction of a million of photoproduced $J/\psi$'s with the $p$ beam and about one thousand with the Pb beam. The latter case has the great advantage that the photon emitter is necessarily the
Pb nucleus. With target densities about 2 orders of magnitude smaller, it seems complicated to perform such a measurement with the Pb beam without storage cell, except for the case of polarised $^3$He$^\uparrow$ for which the injected gas flux can be increased. The latter case is particularly interesting as it allows one to probe polarised neutrons.

We have then used a model of the GPD $E^g$ to predict the magnitue of the STSA. When folded with the expected size of the statistical samples
and the target polarisation, we have found that STSAs can be measured with a precision from 1 to 4 \% for $p$H$^\uparrow$ collisions and 10 to 40 \% for  PbH$^\uparrow$ collisions. In both cases, the accessible range in $x_F$ is from $0$ down to $-0.35$. Overall, we consider these results as a confirmation that the first measurement of the GPD $E^g$ can be made in the fixed-target mode at the LHC by 2025.

Finally, let us emphasise that gaseous deuterium and helium (3 and 4) (un)polarised targets can be used with AFTER@LHCb~\cite{Hadjidakis:2018ifr}. The expected luminosities are at least as large as those discussed here. If one can ensure that the nucleus stays intact, this would provide new means to study the GPDs of these light nuclei~\cite{Berger:2001zb,Cano:2002tn,Guzey:2003jh,Scopetta:2004kj,Mazouz:2007aa,Anikin:2011aa,Rinaldi:2012ft,Dong:2014eya,Dupre:2015jha,Hattawy:2017woc,Fucini:2018gso,Cosyn:2018rdm} .

\section*{Acknowledgements}
We thank S. Klein, J. Nystrand for useful discussions. This work is partly supported 
by the COPIN-IN2P3 Agreement, by the grant 2017/26/M/ST2/01074 of the National Science Center in Poland, by French–Polish scientific agreement POLONIUM.

\bibliographystyle{utphys}
\bibliography{UPC-JPsi-STSA-061218}

\end{document}